# Exciton-Plasmon Coupling Mediated Superior Photoresponse in 2D Hybrid Phototransistors

Shubhrasish Mukherjee[1], Didhiti Bhattacharya[1], Samit Kumar Ray[*1,2] and Atindra Nath Pal[*1]

[1]*S. N. Bose National Center for Basic Science, Sector III, Block JD, Salt Lake, Kolkata – 700106*

[2] *Indian Institute of Technology Kharagpur, 721302, West Bengal, India*

Email: physkr@phy.iitkgp.ac.in, atin@bose.res.in

## Abstract

The possibility of creating heterostructure of two-dimensional (2D) materials has emerged as a viable route towards realizing novel optoelectronic devices. However, the low light absorption due to their small absorption cross section, limits their realistic application. While light-matter interaction mediated by strong exciton-plasmon coupling has been demonstrated to improve absorbance and spontaneous emission in a coupled TMDC and metallic nanostructures, the fabrication of tunable broadband phototransistor with high quantum yield is still a challenging task. By synthesizing Ag nanoparticles (Ag NPs) capped with a thin layer of polyvinylpyrrolidone (PVP) through chemical route, we report a lithography-free fabrication of a large area broadband superior gate-tunable hybrid phototransistor based on monolayer graphene decorated by $WS_2$-Ag NPs in a three-terminal device configuration. The fabricated device exhibits extremely high photoresponsivity (up to $3.2 \times 10^4$ A/W) which is more than 5 times higher than the bare graphene/$WS_2$ hybrid device, along with a low noise equivalent power (NEP) (~$10^{-13}$ W/Hz$^{0.5}$, considering 1/f noise) and high specific detectivity ~$10^{10}$ Jones in the wide (325-730 nm) wavelength region. The additional PVP capping of Ag NPs helps to suppress the direct charge and heat transfer and most importantly, increases the device stability by preventing the degradation of $WS_2$-Ag hybrid system. The enhanced optical properties of the hybrid device are explained via dipole mediated strong exciton-plasmon coupling, corroborated by COMSOL Multiphysics simulation. Our work demonstrates a strategy towards obtaining an environment-friendly, scalable, high-performance broadband phototransistor by tuning the exciton-plasmon coupling for new generation opto-electronic devices.

# Introduction

Two dimensional (2D) materials like graphene, transition metal dichalcognides (TMDCs) attract a lot of attention in the past 15 years because of their atomically thin profile, high transparency, higher carrier mobilities and appealing optoelectronic properties[1,2]. Most importantly, it is possible to create vertical heterostructures by using different 2D materials as they are coupled by the weak Van der Waals force in the out-of-plane direction[3,4] to introduce multiple functionalities. Graphene, a gapless semimetal having higher carrier mobilities, lacks its potential in the optoelectronic applications because of its infirm light absorption (~ 2 % of the incident light)[5,6]. In contrast, TMDCs ($MoS_2$, $WS_2$ etc.) manifest magnificent light-matter interaction characteristics primarily in the UV-Vis region and become the natural partner of graphene[7,8]. Due to the low dimensionality and the reduced dielectric screening, the optoelectronic properties of such 2D materials are influenced by the excitons and trion[9]. Recently, several attempts have been made to improve the optical properties of such TMDCs using the surface plasmon (SP) effects of some novel metals[10,11]. The localized surface plasmon resonance (LSPR) of a metal nanostructure (nanoparticles, nanowires etc.)[12,13] can be excited by an incident electromagnetic radiation and it depends on the carrier density, size and shape of the nanostructures and also on the external surrounding[14,15]. The SP of the metal nanostructures can couple with the excitons of TMDCs and results in an enhanced absorbance, emission and improved photodetection capabilities. On the other hand, formation of van der Waals heterostructure of graphene with photo-active materials like TMDC[7,16], perovskite[17,18,19], semiconducting quantum dots[20,21] provides a promising scheme to design superior photodetector that combine the advantage of light absorption of the photo sensing materials and high charge mobility of graphene as the conducting channel. While there are some reports on TMDC based plasmonic phototransistors demonstrating enhanced performance based on exciton-plasmon coupling, fabrication of tunable hybrid optoelectronic devices with high quantum yield for broadband application is still missing.

In this work, a novel three terminal phototransistor device architecture is reported based on graphene/$WS_2$-Ag nanoparticles heterostructure. The polyvinylpyrrolidone (PVP) capped silver (Ag) nanoparticles (NPs) restricts the direct charge, energy and heat transfer and enhances the optical properties of tungsten disulphide ($WS_2$) by surface plasmon-exciton coupling effect[22]. Using monolayer graphene as a conducting channel and the Ag NPs decorated $WS_2$ nanosheets as the light absorbing material, the device offers superior photodetection capabilities in the broad UV-Vis (325-730 nm) region with an excellent gate tunability. The fabricated highly stable graphene/$WS_2$-Ag phototransistor exhibits a very high photoresponsivity (R) > $10^4$ A/W, higher specific detectivity ($D^*$) > $10^{10}$ Jones and low noise equivalent power (NEP) ~ $10^{-13}$ W/$Hz^{0.5}$ in the overall spectrum region which is several orders of magnitude higher than the bare graphene/$WS_2$ control phototransistor. By tuning the exciton-plasmon coupling, the demonstrated stable, sensitive phototransistor devices have the potential for the next generation optoelectronic device applications.

## Result and discussions

**Figure 1a** represents the typical TEM (transmission electron microscopy) image of chemically exfoliated $WS_2$ decorated with Ag nanoparticles (NPs). The spherical Ag nanoparticles are encapsulated with a thin insulating polyvinylpyrrolidone (PVP) layer (TEM images and Selected Area Electron Diffraction (SAED) pattern of bare $WS_2$ and Ag NPs are presented in **Figure S1a**, **b** and **c**). The crystalline states of Ag provide a lattice spacing ~0.23nm, corresponds to (111) plane[23] of Ag, and the same for $WS_2$ is ~0.27 nm, indicating (100) plane[24] of $WS_2$, as evident from the high-resolution TEM (HRTEM) image (**Figure 1b**). The SAED pattern of Ag NPs decorated $WS_2$ nanosheets exhibits distinct diffraction spots which can be assigned to (100), (110) planes of $WS_2$ and (111) planes[25] of Ag (**Figure 1c**). This PVP based encapsulation of the Ag nanoparticles has multiple advantages. Firstly, it helps to supress the direct charge, energy transfer and the band gap pinning[26] and hence, the Ag nanoparticles can interact with $WS_2$ through localized surface plasmon resonance (LSPR) effect. Secondly, it improves the stability of Ag nanoparticles by stopping the chemical reaction between Ag and $WS_2$, which was reported to form $Ag_2S$ even in the room temperature[27]. The crystallinity of bare $WS_2$ and Ag coupled $WS_2$ nanosheets have been examined by X-ray diffraction (XRD). The formation of (002), (004) and (006) planes is the signature of crystalline $WS_2$ layers[24] and the appearance of (111) plane signifies the presence of Ag in the $WS_2$-Ag hybrid system (**Figure 1d**). The Atomic Force Microscopy (AFM) images of bare $WS_2$ and $WS_2$-Ag hybrid suggest the thickness of typical nanosheets are ~ 2/3 nm (**Figure S1d** and **S1e**). The Raman spectra of few layered $WS_2$ sheets and $WS_2$-Ag hybrid are represented in **Figure 1e**. The in-plane ($E^1_{2g}$) and out-of-plane ($A_{1g}$) Raman modes are observed at 352 $cm^{-1}$ and 419 $cm^{-1}$ respectively for bare $WS_2$ suggesting the $WS_2$ are few layered thick[28]. The intensities of both the peaks are found to be enhanced significantly because of the LSPR enhancement by Ag NPs decorated on WS$_2$ (**Figure S2a**)[29]. Additionally, the in-plane E$^1_{2g}$ mode of WS$_2$ is red shifted due to the presence of Ag NPs as shown in **Figure 1e**. The red shift of E$^1_{2g}$ mode can be ascribed to the strain relaxation due to lattice mismatch between Ag and WS$_2$[30,31]. The out-of-plane A$_{1g}$ mode behaves oppositely in the hybrid WS$_2$-Ag system. The strongly localized electromagnetic field of the plasmonic Ag NPs can stiffen the vertical oscillations of S atoms in WS$_2$, resulting in a blue shift of A$_{1g}$ mode[32]. With increasing the Ag NPs concentration, the peak position of A$_{1g}$ mode is blue- shifted monotonically, while the red-shifting of E$^1_{2g}$ mode saturates after a certain Ag:WS$_2$ ratio (**Figure S2b**). The higher Ag concentration helps to increase the exciton-plasmon coupling (due to the strong surface plasmon of Ag) which results a continuous blue shift of A$_{1g}$ peak. The saturation of E$^1_{2g}$ mode shift is associated with the saturation of the strain effect on WS$_2$ layers due to the presence of Ag NPs on top.

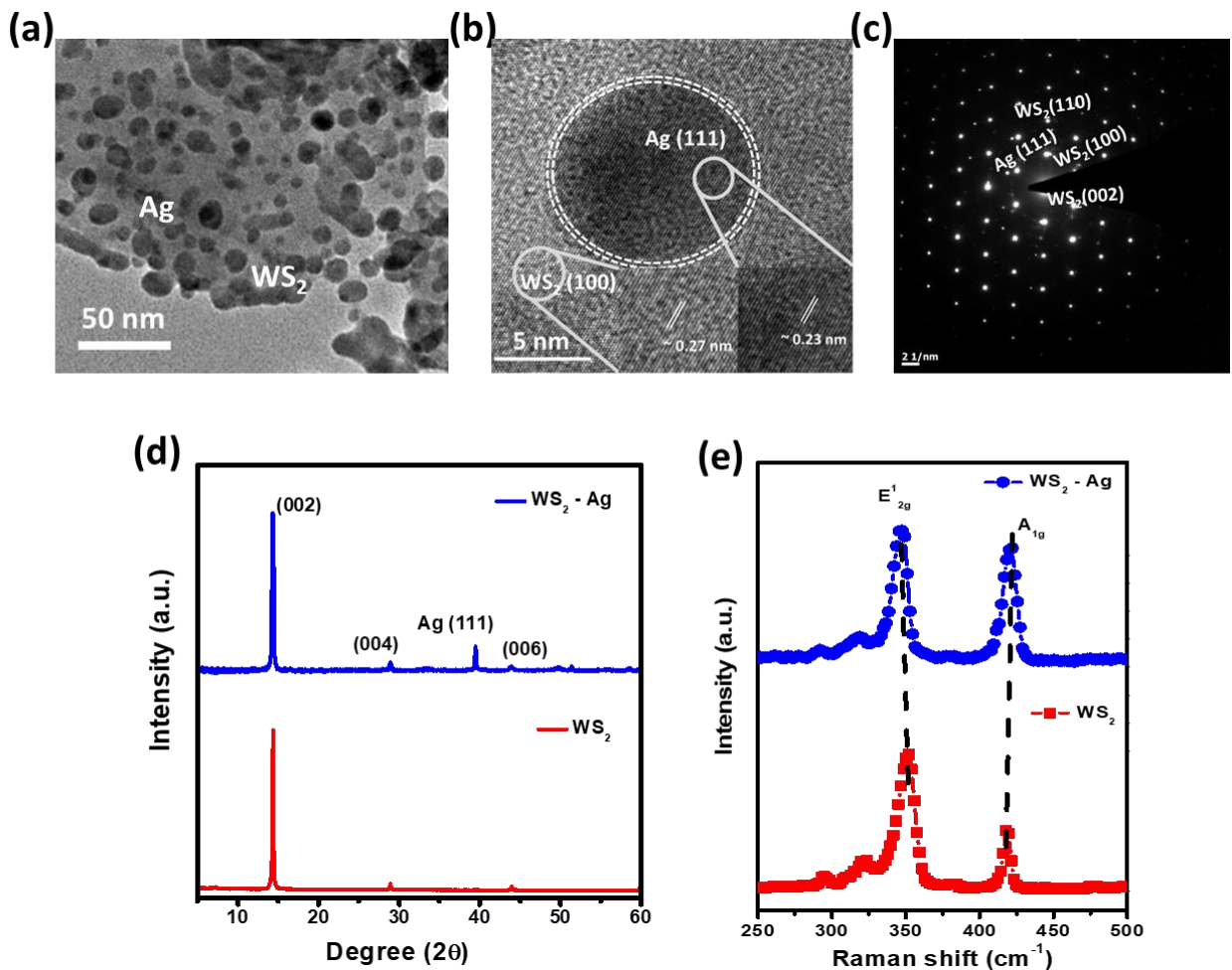

**Figure 1:** Structural and Raman characterisations of bare WS$_2$ and WS$_2$-Ag hybrid structure. (a) Transmission electron microscopy (TEM) image of Ag NPs decorated WS$_2$ layers. (b)

HRTEM image of the hybrid indicating different lattice planes (inset) such as (100) planes of $WS_2$ and (111) planes of Ag NPs. (c) SAED pattern of the hybrid. (d) XRD patterns of $WS_2$ and the $WS_2$-Ag hybrid. (e) Raman spectra of bare $WS_2$ and the hybrid showing the characteristic in plane $E^1_{2g}$ and out of plane $A_{1g}$ peak (under 532 nm excitation).

UV-Vis absorption spectra of bare $WS_2$, Ag NPs and $WS_2$-Ag hybrid are depicted in **Figure 2a**. The characteristics exciton 'A' and 'B' peaks of $WS_2$ remain intact in $WS_2$-Ag hybrid system[33]. The overall absorbance of $WS_2$ is enhanced due to the LSPR induced local electromagnetic interaction between the Ag NPs and semiconducting $WS_2$ [34,35]. Excitonic A peak of $WS_2$ can be more clearly visualized in the second derivative of the absorption spectra (**Figure S2c**). A significant blue shift is observed in the hybrid $WS_2$-Ag suggesting the impact of exciton-plasmon coupling that supports the observation of the Raman spectra[36].

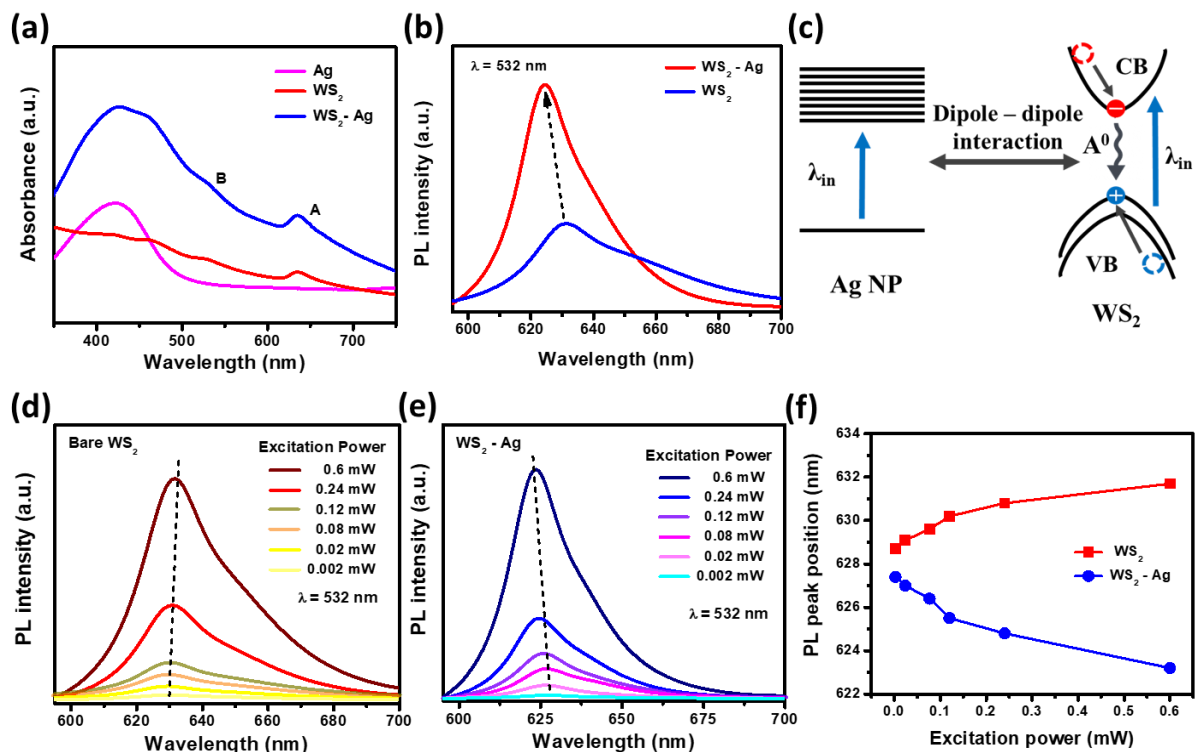

**Figure 2:** Analysis of the spectroscopic characterisations of bare $WS_2$ and $WS_2$-Ag hybrid structure. (a) Absorbance spectra of bare $WS_2$, Ag nanoparticles (NPs) and the plasmonic hybrid. Two distinct exciton peaks (exciton 'A' and 'B') are present in $WS_2$ and $WS_2$-Ag hybrid system. (b) Photoluminescence (PL) spectra of bare $WS_2$ and the $WS_2$-Ag hybrid structure. (c) Schematic diagram of the dipole mediated exciton-plasmon coupling in the hybrid structure. (d) and (e) power dependent PL spectra of control $WS_2$ layer and the $WS_2$-Ag hybrid structure

respectively. (f) The PL peak position as a function of excitation power (λ = 532 nm) for both WS$_2$ and WS$_2$-Ag hybrid.

The role of surface plasmon resonance of the Ag NPs can be well understood from the concentration dependent absorption spectra in WS$_2$-Ag hybrid structure. Interestingly it is observed that, with increasing Ag NPs density, the exciton to plasmon peak ratio decreases whereas the exciton 'A' peak intensity increases (**Figure S2d**), suggesting the possibilities to tune the coupling strength (exciton-plasmon) by controlling the concentration ratio of the materials. The effect of SPR on the optical properties of WS$_2$ is further studied through micro-Photoluminescence (PL) spectroscopy (excitation 532 nm). Having a weak asymmetric profile, the PL spectra of WS$_2$ can be deconvoluted into three possible components such as neutral exciton ($A^0$), trion ($A^T$) and biexciton ($A^A$) (**Figure S3a**)[37,38]. The PL emission spectra is dramatically enhanced in the hybrid WS$_2$-Ag system as shown in **Figure 2b**. The observed PL enhancement from the WS$_2$-Ag hybrid is associated with the combined contribution of the enhanced light absorption by WS$_2$ in the hybrid structure, the exciton-plasmon-photon conversion and plasmon enhanced exciton generation-recombination process[39,40]. The insulating PVP layer prevents the PL quenching, band-pinning and doping by avoiding the direct metal-semiconductor contact[41,42]. The excitons in WS$_2$ can couple with the surface plasmon (SP) of Ag NPs due to their spatial wave-functional overlap and resulting an enhanced emission[43]. Compared to the other excitons ($A^T$ and $A^A$), only the intensity of neutral exciton ($A^0$) dominates in the hybrid WS$_2$-Ag structure as shown in **Figure S3b**. The realization of exciton-plasmon coupling requires simultaneous spectral and spatial overlap between excitons and plasmons[44]. Thus, the confined excitons are spectrally coupled with the plasmon, with consequent strong transition dipole moments interactions. The dipole-dipole interaction between the exciton states and the SP is depicted schematically in **Figure 2c**. This dipolar coupling rate is dependent on the oscillator strength of excitons and the local field enhancement of SP[45,46]. In order to get deeper insight of the exciton-plasmon coupling mechanism, the excitation power dependent PL emission spectra are represented in **Figure 2d** and **2e** for control WS$_2$ and WS$_2$-Ag hybrid respectively. It is observed that, the PL peak position of WS$_2$-Ag hybrid shifts towards shorter wavelength with increasing excitation power while the bare WS$_2$ shows an opposite trend. With increasing excitation power, the emission spectra of WS$_2$ exhibits a continuous red shift (**Figure 2d**), as all three excitons ($A^0$, $A^T$ and $A^A$) contributed in the emission spectra[38]. On the other hand, the contribution of major PL emission of WS$_2$-Ag hybrid is dominated by neutral exciton ($A^0$) only as discussed previously. The reduced band filling effect due to the dipolar interaction between the excitons of WS$_2$ with SP of Ag NPs results in a continuous blue shift with increasing excitation power in WS$_2$-Ag hybrid[12]. The PL peaks shift with excitation power for both bare WS$_2$ and WS$_2$-Ag is depicted in **Figure 2f**. Both bare WS$_2$ and WS$_2$-Ag hybrid exhibit a monotonic increase in PL intensity as a function of illumination power (for 532 nm) which is represented in **Figure S3c**. To estimate the lifetime of the photogenerated charge carriers, time resolved photoluminescence (TRPL) measurements are carried out for bare WS$_2$ and WS$_2$-Ag hybrid under identical conditions (excitation: 409 nm and emission at 500 nm) (**Figure S3d**). It is observed that the life time of bare WS$_2$ (1.67 ns) slightly decreases when coupled with Ag NPs. The quenched lifetime in WS$_2$-Ag hybrid (1.53 ns) indicates the enhancement of spontaneous emission rate of the hybrid system due to the enhanced exciton plasmon coupling. From the steady state PL, it is observed that the neutral exciton dominates in WS$_2$-Ag system and the spontaneous emission rate of neutral exciton is amplified by the dipole-dipole interactions (Purcell effect)[15].

The hybrid phototransistor devices have been fabricated by considering CVD grown graphene as the conducting channel between two Ti/Au electrodes and PVP capped Ag nanoparticle decorated WS$_2$ (or bare WS$_2$) as the photo absorbing layer. **Figure 3a** represents the schematic of the three-terminal device, which is basically a graphene transistor on a Si/SiO$_2$ substrate sensitized by WS$_2$-Ag hybrid on top of it. The photoresponse characteristics (I$_{ph}$-V$_{ds}$) of the graphene/WS$_2$-Ag and graphene/WS$_2$ devices with different illumination wavelengths (at a constant illumination intensity ~ 10 μW/cm$^2$, V$_{bg}$ = 0 V) are shown in the **Figure 3b** and **3c** respectively. Here, it is found that, the photocurrent (I$_{ph}$) increases linearly with the increase of source-drain (V$_{ds}$) voltage varying from 0 to 1 V for both the devices[16]. **Figure 3d** shows the spectral photoresponsivity (R) of the hybrid devices with a constant V$_{ds}$ = 1 V, back-gate voltage (V$_{bg}$) of 0 V and illumination power density (P$_{LED}$) ~ 1 μW/cm$^2$ where, R can be calculated as[47]

$$R = \frac{I_{ph}}{P_{eff}} \quad \ldots\ldots\ldots\ldots\ldots\ldots\ldots\ldots\ldots\ldots\ldots\ldots\ldots \quad (1)$$

Here, I$_{ph}$ = (I$_{light}$ - I$_{dark}$) is the photocurrent and P$_{eff}$ is the effective optical illumination power on the surface of the device.

Furthermore, the photoresponsivity (R) is noticeably enhanced for the plasmonic device (graphene/WS$_2$-Ag) in the whole UV- Vis spectrum region (325-730 nm) compared to the graphene/WS$_2$ control device. The enhancement of R is the highest (~5.2 times) at 410 nm of illumination. This enhanced responsivity can be attributed to the localized surface plasmon resonance effect of Ag nanoparticles which enhances the optical absorbance of the WS$_2$ layers.

Additionally, the strong exciton-plasmon coupling of $WS_2$-Ag enhances the photoresponsivity significantly (~ 4.5 times) at 633 nm of illumination.

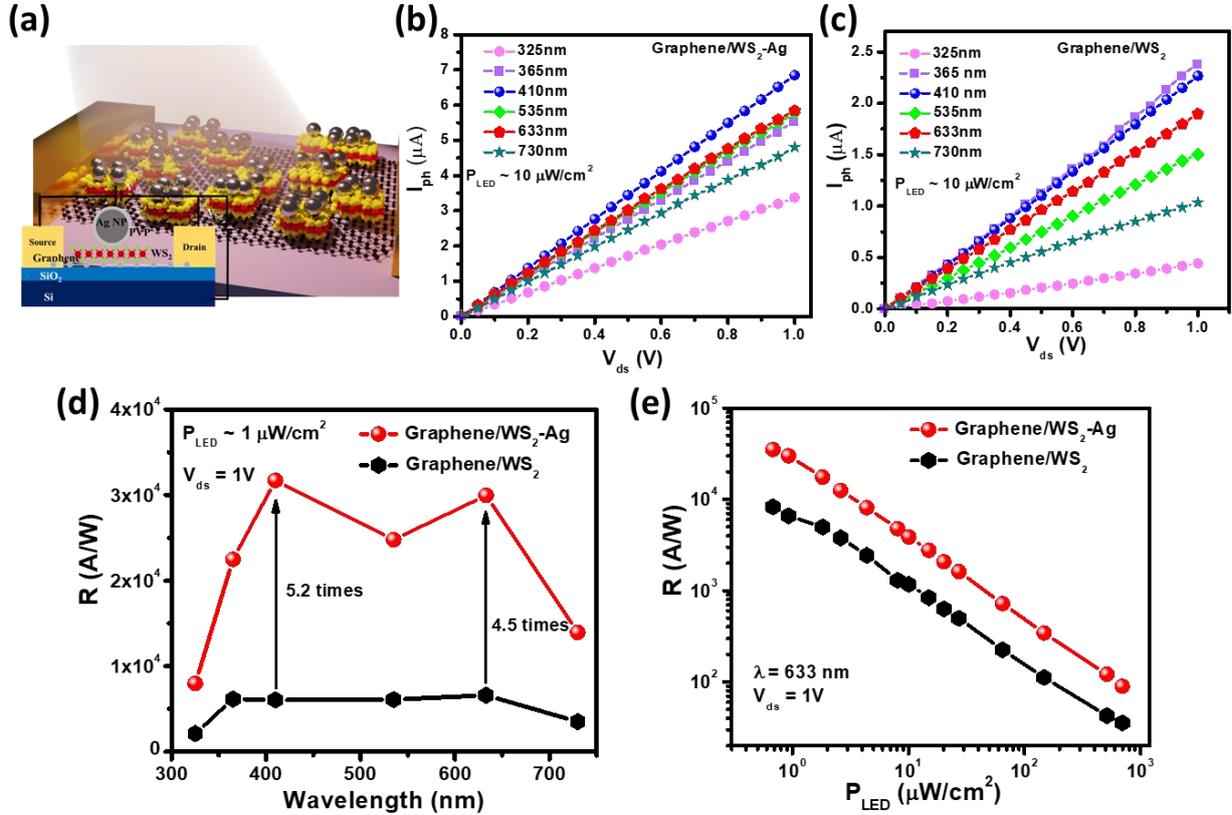

**Figure 3:** Comparison of the photoresponse between the hybrid devices (a) Schematic of the hybrid graphene/$WS_2$-Ag phototransistor. Photocurrent ($I_{ph}$) as a function of applied source-drain bias ($V_{ds}$) voltage of (b) graphene/$WS_2$-Ag and (c) graphene/$WS_2$ devices. Photoresponsivity as a function of (d) wavelength and (e) illumination power ($P_{LED}$) of the plasmonic hybrid and control devices.

**Figure 3e** represents a comparison of the photoresponsivity as a function optical illumination power density ($P_{LED}$) of the graphene/$WS_2$-Ag and graphene/$WS_2$ devices with $\lambda$ = 633 nm, $V_{ds}$ = 1 V for zero gate voltage ($V_{bg}$ = 0 V). At a lowest illumination power of ~ 0.2 $\mu W/cm^2$ the plasmonic graphene/$WS_2$-Ag device offers the highest photoresponsivity (R) of ~ $3.5 \times 10^4$ A/W, whereas, in case of non-plasmonic control device, R becomes ~ $8.2 \times 10^3$ A/W at 633 nm. For both the devices, R decreases monotonically with the increase of optical power. The saturation of photocurrent ($I_{ph}$) due to lowering of the interfacial built-in-field reduces the responsivity with the increase of illumination power[48,49].

Noise equivalent power (NEP) and specific detectivity (D*) are two important parameters to evaluate the capability of weak signal detection of a photodetector. The NEP of a photodetector is defined as[47]

$$NEP = \frac{S_I}{R} \quad \text{……………………………..} \quad (2)$$

Where, $S_I$ id the total noise current considering 1/f noise, shot noise and the thermal noise of the devices and R represents the photoresponsivity.

Similarly, the specific detectivity (D*) is defined as[47]

$$D^* = \frac{\sqrt{A}}{NEP} \quad \text{……………………………..} \quad (3)$$

where, A is the area of the device.

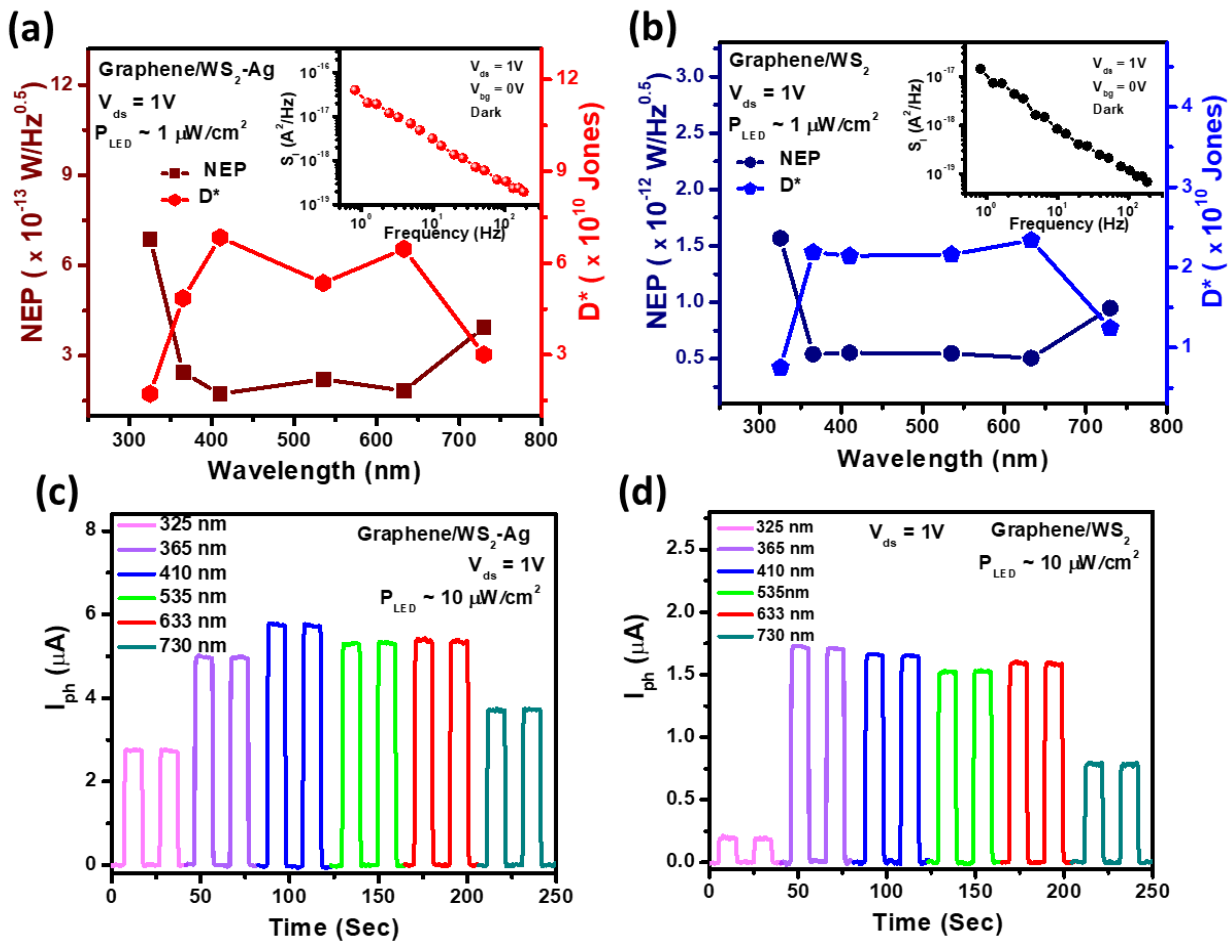

**Figure 4:** Noise equivalent power (NEP) and the specific detectivity (D*) of (a) hybrid graphene/WS$_2$-Ag and (b) graphene/WS$_2$ devices. 1/f noise spectra of the respective devices are shown in the inset. (c) and (d) The temporal photoresponse in the broadband UV-Vis (325-730 nm) region of the graphene/WS$_2$-Ag and graphene/WS$_2$ hybrid devices respectively.

At a frequency of 1 Hz, the measured 1/f noise spectral density ($S_I$ (1/f)) of the plasmonic graphene/WS$_2$-Ag and graphene/WS$_2$ devices are calculated to be $3.01 \times 10^{-17}$ A$^2$/Hz and $1.11 \times 10^{-17}$ A$^2$/Hz, respectively. This 1/f noise (inset **Figure 4a** and **4b**) clearly dominates over the other source of noises in these hybrid phototransistor devices (See **supporting information note 4** for the calculation of the total noise spectral density). **Figures 4a** and **4b** show the spectral NEP and the specific detectivity (D$^*$) of the hybrid devices. It is seen that the plasmonic graphene/WS$_2$-Ag device offers lower NEP and higher specific detectivity as compared to the bare graphene/WS$_2$ device in overall spectrum region. For example, the calculated NEP becomes $1.73 \times 10^{-13}$ W/ Hz$^{0.5}$ and $5.52 \times 10^{-13}$ W/ Hz$^{0.5}$ and the specific detectivity (D$^*$) becomes $6.83 \times 10^{10}$ Jones and $2.14 \times 10^{10}$ Jones for the plasmonic graphene/WS$_2$-Ag and bare graphene/WS$_2$ devices, respectively (at λ = 410 nm, V$_{ds}$ = 1 V). A lower NEP and higher detectivity indicate the superiority of the plasmon coupled device for weak light detection.

**Figures 4c** and **4d** represent the UV-Vis (325-730 nm) broadband photo-switching characteristics of these two phototransistors with the same experimental conditions V$_{ds}$ = 1 V, V$_{bg}$ = 0 V and P$_{LED}$ ~ 10 μW/ cm$^2$. Both the devices show stable and repeatable temporal photoresponse by following the multiple ON/OFF illumination cycles. The plasmonic graphene/WS$_2$-Ag device offers a significantly higher photocurrent with consistent reproducibility in the overall wavelength region compared to the bare graphene/WS$_2$ device. This higher temporal photocurrent is consistent with the enhanced photoresponsivity and can be described by the strong exciton-plasmon coupling as explained before. The characteristic temporal response time is a very important parameter to evaluate the performance of a photodetector. The rise time (τ$_{Rise}$) and decay time (τ$_{Fall}$) of a photodetector (**Figure S4a** and **4b**) device are basically defined as the time gap for the current changes from 10% to 90% and vice-versa when light is turn ON or OFF[50]. The rise times (τ$_{Rise}$) of the graphene/WS$_2$ phototransistor with and without Ag nanoparticles are 0.49 sec and 0.52 sec, respectively, while the corresponding fall times (τ$_{Fall}$) are 0.50 sec and 0.57 sec. The transfer characteristic (I$_{ds}$-V$_{bg}$) of bare graphene and WS$_2$-Ag (PVP) decorated graphene transistor in dark is represented in **Figure S5a**. No charge neutrality point (V$_D$) is observed in the pristine graphene device, mostly due to the hole-doping effect caused by the PMMA- based wet transfer of the graphene film[51]. The transfer characteristics do not get altered due to the deposition of WS$_2$-Ag (PVP), except the appearance of the V$_D$ at ~ 23 V. This shift of V$_D$ implies the transfer of electrons from Ag decorated WS$_2$ to graphene as shown schematically in **Figure S5b**. This charge transfer between WS$_2$ and graphene layers assists to equilibrate the Fermi levels and consequently the valance band (VB) and the conduction band (CB) of WS$_2$ bend upward at graphene/WS$_2$ interface.

**Figure 5a** shows the transfer characteristics of the hybrid device before and after illuminating a 633 nm radiation (left axis) and the gate tunable photoresponsivity at V$_{ds}$ = 0.5 V (right axis). It is seen that the photoresponsivity can be tuned significantly by the application of a gate voltage (V$_{bg}$). At 633 nm of illumination with 8 µW/cm$^2$ power, the photoresponsivity increases from 2.57 × 10$^3$ A/W to 3.32 × 10$^3$ A/W when the gate voltages (V$_{bg}$) changes from -15 V to 0 V, while photoresponsivity (R) again decreases to -3.06 × 10$^2$ A/W at V$_{bg}$ = 45 V.

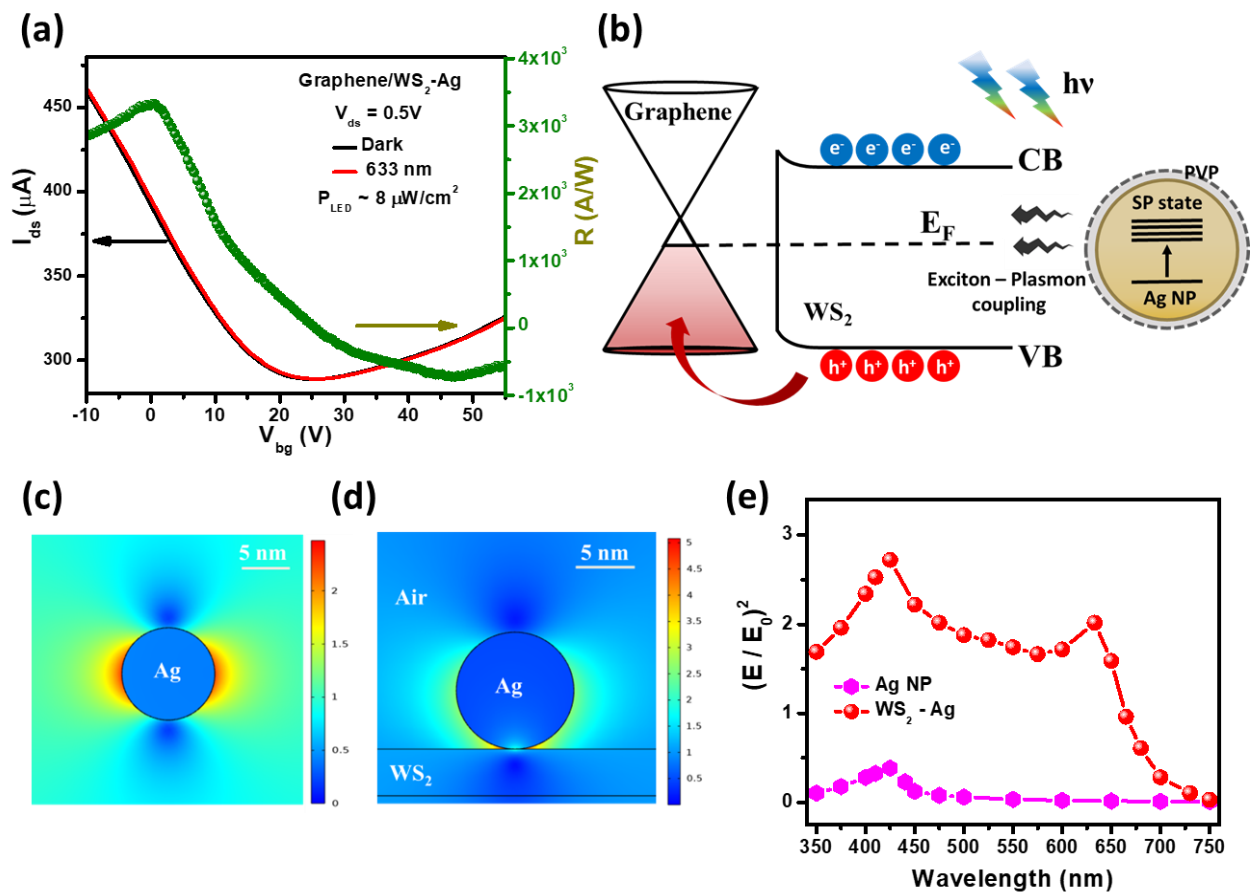

**Figure 5:** Photodetection characteristics of graphene/WS$_2$-Ag heterostructure. (a) Transfer characteristics and the gate tunable photoresponse of the plasmonic hybrid device at λ = 633 nm, V$_{ds}$ = 0.5V, P$_{LED}$ ~ 8 µW/cm$^2$. (b) Suggested energy band diagram illustrating the gate tunable photoresponse in the hybrid device. (c) and (d) Electric field distribution of a single Ag NP and the WS$_2$-Ag interface under the excitation of 420 nm. (e) The simulated spectral electric field intensity of bare Ag NP and WS$_2$-Ag hybrid structure.

Such a polarity tunable photoresponsivity can be described by a simple energy band diagram as shown in **Figure 5b**. Upon irradiation, electron-hole pairs are generated in Ag NPs decorated

WS$_2$. The photogenerated holes are then transferred to graphene channel by trapping the electrons in WS$_2$ due to the upward band bending. While decreasing the gate voltage to a value of V$_{bg}$ < V$_D$, the Fermi level of the graphene is lowered and the graphene becomes hole doped. It helps to transfer more photogenerated holes to the graphene channel by enhancing the interfacial electric field. As a result, the photoresponsivity of the device increases. Similarly, when the applied gate voltages are higher than the Dirac voltage (V$_{bg}$ > V$_D$), the graphene becomes electron doped. Now, increasing V$_{bg}$ also increases the injection of photogenerated electrons from WS$_2$ to graphene due to downward band bending in WS$_2$, resulting in a negative photoresponsivity. Interestingly, the photoresponsivity becomes negligibly small near the Dirac point (V$_D$ ~ 23 V) making the gate voltages as an additional switch to tune the photoresponsivity of the devices. The presence of an additional PVP layer in between WS$_2$ and Ag NPs effectively blocks all the transfer mechanism (charge, heat etc.) except the LSPR coupling. The dipole-dipole energy transfer effectively increases the electric field intensity at the locality of semiconducting WS$_2$ and results in an enhanced optical absorption accompanied by the generation of larger density of electron-hole pairs upon illumination. To understand the effect of surface plasmon (SP) coupling for synergistically enhanced absorption, emission and improved photoresponse in the WS$_2$-Ag hybrid, the electric field distribution is investigated by using COMSOL Multiphysics simulation (see **Experimental section** for details). Under resonance excitation ~ 420 nm, the simulated electric field distribution of a single Ag NP and WS$_2$-Ag hybrid system are represented in **Figure 5c** and **5d,** respectively. This strong, tightly confined plasmonic mode of Ag NPs is responsible for the enhanced light matter interaction in the hybrid system. The excitation rate and the absorption cross-section of the excitonic system are enhanced by a factor ($\left|E/E_0\right|^2$), where E and E$_0$ are the local and incident electric fields respectively[52,53]. The simulated spectral distribution (350-750 nm) of the interfacial electric field is depicted in **Figure 5e** for a single Ag NP and WS$_2$-Ag hybrid system. The field enhances in the hybrid system for the overall spectrum region, which is consistent with the absorption spectra (**Figure 2a**). This enhancement of the interfacial electric field can be qualitatively explained by a simple image charge model. The electromagnetic field of the plasmonic Ag NP can effectively be screened due to the introduction of the WS$_2$ layer in the vicinity and therefore generates the image charges. The coupling between the plasmonic particle and its image charge pushes down the hot spot of the generated electric field and subsequently enhances the interfacial field of the hybrid WS$_2$-Ag system.

Finally, the stability of the plasmonic hybrid photodetector is checked by keeping it in the normal ambient conditions (in a vacuum desiccator). The device offers extremely stable photocurrent (**Figure S6**) even after 3 months of its fabrication. The insulating PVP capping therefore greatly helps to stabilize the device by preventing the direct contact of such Ag NPs with the $WS_2$ layers. Notably, highly enhanced photoresponsivity, along with good stability and durability, make this hybrid graphene/$WS_2$-Ag phototransistor superior to other reported results (presented in **Table-1**) and very promising for the future multifunctional optical devices.

## **Summary**

In summary, we have demonstrated a stable, cost effective, scalable and gate tunable phototransistor based on monolayer graphene/$WS_2$-Ag NPs plasmonic heterostructure. The presence of strong surface plasmon resonance due to synthesized Ag NPs effectively enhances the light matter interactions in $WS_2$. The enhanced optical properties in the $WS_2$-Ag hybrid are originated from the dipole mediated exciton-plasmon coupling in the multiple integrated hot spots. Compared to the bare graphene/$WS_2$ device, the graphene/plasmonic Ag NPs coupled $WS_2$ hybrid device offers superior photodetection capabilities in the broadband UV-Vis (325-730 nm) region with an excellent gate tunability. Furthermore, it shows the photoresponsivity as high as ~ $3.2 \times 10^4$ A/W and ~ $2.9 \times 10^4$ A/W for 410 nm and 633 nm of illumination, respectively which are about 500% higher than the control graphene/$WS_2$ device. Also, considering 1/f noise, the plasmon coupled graphene/$WS_2$ phototransistor offers a very low noise equivalent power (NEP) ~$10^{-13}$ W/$Hz^{0.5}$ and high specific detectivity (D*) ~$10^{10}$ Jones in the wide (325-730 nm) wavelength region. The additional PVP capping in Ag NPs helps to supress the loss through the heat and energy transfer and most importantly increases the device stability by preventing any degradation. The strong exciton-plasmon coupling mediated enhanced optical properties of the plasmonic $WS_2$-Ag hybrid have been explained by COMSOL Multiphysics simulation. Our study provides a strategic route towards facile fabrication of a superior quality large area, broadband, plasmon based, highly stable hybrid phototransistor based on the exciton-plasmon coupling, a potential candidate for the next generation optoelectronic devices.

**Table 1. Performance comparison of some graphene based plasmonic photodetectors:**

| SL No | Device Structure | Wavelength (Nm) | $V_{ds}$ (V) | Responsivity (R) A/W | Rise time (sec) | Decay time (sec) | Ref |
|---|---|---|---|---|---|---|---|
| 1. | Graphene – perovskite – Au nanoarray | 375 – 808 | 2 | 18.71 | 0.33 | 0.27 | 17 |
| 2. | Graphene – MAPbI3 – Au NPs | 532 | 10 | $2.1 \times 10^3$ | 1.5 | 1.5 | 18 |
| 3. | Graphene – $WS_2$ – Ag NPs | 400 – 750 | 0.8 | 11.4 | 0.3 | 1 | 35 |
| 4. | Graphene– MAPbI3– Au nano stars | 532 | -5 | $5.9 \times 10^4$ | 2.5 | 11.9 | 19 |
| 5. | Graphene – Ag NP | 225 – 450 | 0.5 | 14.5 | 6 | 17 | 50 |
| 6. | Graphene – Ag NP | 250 – 450 | 0.35 | 82 | 1 | 4 | 52 |
| 7. | Graphene – perovskite – Au NPs | 450 – 800 | 0.1 | 495.3 | 7 | 7 | 49 |
| 8. | Graphene – $WS_2$ – Ag (PVP) | 325 – 730 | 1 | $3.5 \times 10^4$ | 0.49 | 0.52 | **This work** |

# **Experimental Section**

### **Chemicals:**

All chemicals (Tungsten disulphide, Lithium Bromide, Isopropyl Alcohol, Polyvinylpyrrolidone, Silver Nitrate, Sodium Citrate) are brought from Sigma-Aldrich without any further purification.

### **Synthesis:**

In this work, we have reported the unique properties of highly stable, solution-processed, plasmonic Ag-nanoparticles decorated and chemically exfoliated $WS_2$ dispersion, in this regard layered WS$_2$ nanosheets are synthesized from bulk WS$_2$ by Li-intercalation assisted chemical exfoliation technique and Polyvinylpyrrolidone (PVP) coated silver nanoparticles are added in the dispersion of WS$_2$ to prepare WS$_2$-PVP encapsulated Ag nanoparticle nanocomposite. WS$_2$ nanosheets are synthesized chemically by using Li-intercalation technique. First, for exfoliation of WS$_2$ nanosheets, bulk WS$_2$ powder 2.5 gm with anhydrous LiBr at 1:1 molar ratio is dispersed in 25ml hexane solution. This solution is further sonicated for 6 hrs. by using bath sonicator. After sonication the resulting black dispersion is centrifuged at 5000 rpm for 15 mins to remove hexane and untreated Li ions. Then the wet sediments are washed by dispersing in IPA by shaking followed by centrifugation (5000 rpm, 15 mins). By repeating this procedure three times the wet sediment of WS$_2$ is completely transferred in IPA solvent. After 2 hrs. of bath sonication, the resulting dispersion is centrifuged at 5000 rpm for 10 mins and a greenish colour exfoliated WS$_2$ nanosheets are obtained. To prepare silver nanoparticle, PVP is dissolved in DI water and is kept for stirring at room temperature and then silver nitrate (AgNO$_3$) is added in this dispersion. After that Sodium Citrate (NaC) is used as reducing reagent, added dropwise in AgNO$_3$ dispersion and the colour of the solution is turned light orange which indicates the formation of silver nanoparticle. Thereafter different quantity silver nanoparticles are mixed with chemically exfoliated WS$_2$ nanosheets and kept stirring for 1 hr to get the homogeneous mixture of WS$_2$-PVP encapsulated Ag nanoparticle nanocomposite without any precipitation.

**Characterizations:**

The phase and crystallinity of synthesized WS$_2$ nanosheets and WS$_2$-Ag heterostructures are investigated by X-ray diffraction (PANalytical X-PERT PRO) using Cu-K$\alpha$ radiation (1.54 A°). Surface profile of all the samples is examined by using an optical microscope and also in a field-emission scanning electron microscope (FE-SEM) with an electron energy of 20 keV and equipped with an energy-dispersive X-ray (E-DAX) spectrometer. To get the deeper understanding about the morphology, the synthesized samples are investigated using a high-resolution transmission electron microscope (FEI-TECNAI G2 20ST, energy 200 keV) and atomic force microscopy (di INNOVA). Absorption spectrum are measured using UV-Vis spectrometer (Shimadzu -UV-Vis 2600 Spectrophotometer) and Raman and Photoluminescence spectrum are carried out in LabRam HR Evolution; HORIBA France SAS-532nm laser. Time-resolved emission transients are recorded using a time correlated single photon counting (TCSPC) setup using 409 nm excitation. The LED powers are calibrated by using Flame-Ocean Optics spectrometer with integrating sphere set up.

**Device Fabrication:**

CVD-grown monolayer graphene on $p^+$ doped Si/SiO$_2$ (300 nm) substrates (Graphenea, USA) is used to fabricate the phototransistor devices. The electrodes are deposited through a shadow mask by electron beam evaporation to make a channel of W/L = 200 µm/70 µm. P$^+$ doped Si acts as the back gate of the fabricated devices. After that, chemically exfoliated WS$_2$ and plasmonic WS$_2$-Ag (PVP) are simply spin coated on top of the graphene channel to make the complete hybrid devices. Finally, all the phototransistor devices are annealed at 80º C for 1 hr to improve the hybrid interfaces.

**Device Characterisations:**

A homemade electronic setup with an optical window is used for the electrical and the optical experiments. Keithley 2450 sourcemeter and MFLI lock in amplifier (Zurich Instruments) are used in AC 2 probe configuration with a carrier frequency of 226.6 Hz. Collimated and well calibrated Thorlab LEDs powered by DC2200 power supply are used for the photocurrent measurements. The LED powers are calibrated by using Flame-Ocean Optics spectrometer with integrating sphere set up. The whole experiments are performed in room temperature and in vacuum 10$^{-5}$ mbar.

**Numerical Simulation:**

The plasmonic coupling effect on this TMDC-metal nanoparticles (NPs) hybrid system is investigated by optical simulations using COMSOL Multiphysics 5.5 (wave optics module) software. In this module, the incident Gaussian electromagnetic wave (plane wave) is considered to be polarized along x axis and travelling along y axis with normalized amplitude 1V/m. The light matter interaction of the hybrid system is studied by simulating an Ag NP with 10 nm diameter sitting on top of a 3 nm thick WS$_2$ layer. The propagation parameter ($K = \frac{2\pi}{\lambda}$) satisfies electromagnetic wave equation

$$\nabla^2 E = \mu\varepsilon \frac{\delta^2 E}{\delta t^2}$$

Where, μ is the permeability (taken to be 1) and ε is the complex permittivity of the medium which is taken from the previous literatures[54,55]. The interfacial electric field as a function of wavelength is plotted and the extracted values give information regarding the field enhancement with broadband spectral coverage.

## Acknowledgements

S.M. acknowledges the INSPIRE Fellowship program, DST, Govt. of India, for providing him with a research fellowship (IF170929). The authors acknowledge the characterization facilities from the TRC project, clean room fabrication facilities of SNBNCBS and discussion with Rajib Kumar Mitra.

## Conflict of Interest

The authors declare no conflict of interest.

# Supporting information

# Exciton-Plasmon Coupling Mediated Strong Photoresponse in 2D Hybrid Phototransistors

Shubhrasish Mukherjee[1], Didhiti Bhattacharya[1], Samit Kumar Ray[*1,2] and Atindra Nath Pal[*1]

[1]*S. N. Bose National Center for Basic Science, Sector III, Block JD, Salt Lake, Kolkata – 700106*

[2] *Indian Institute of Technology Kharagpur, 721302, West Bengal, India*

Email: physkr@phy.iitkgp.ac.in, atin@bose.res.in

**Contents**

**Supplementary Note 1:** Microstructural characterisation of $WS_2$, $WS_2$-Ag hybrid and Ag nanoparticles (NPs):

**Supplementary Note 2:** Spectroscopic characterisations of $WS_2$ and the plasmonic $WS_2$-Ag hybrid system:

**Supplementary Note 3:** Photoluminescence analysis of $WS_2$ and the plasmonic $WS_2$-Ag hybrid system:

**Supplementary Note 4:** Calculations of the spectral density of noise of the hybrid phototransistors:

**Supplementary Note 5:** Comparison of temporal photoresponse characteristics of the hybrid devices:

**Supplementary Note 6:** Transfer characteristics and the suggested energy band diagrams:

**Supplementary Note 7:** Stability of the hybrid graphene/$WS_2$-Ag phototransistor:

# 1. Microstructural characterisation of WS$_2$, WS$_2$-Ag hybrid and Ag nanoparticles (NPs):

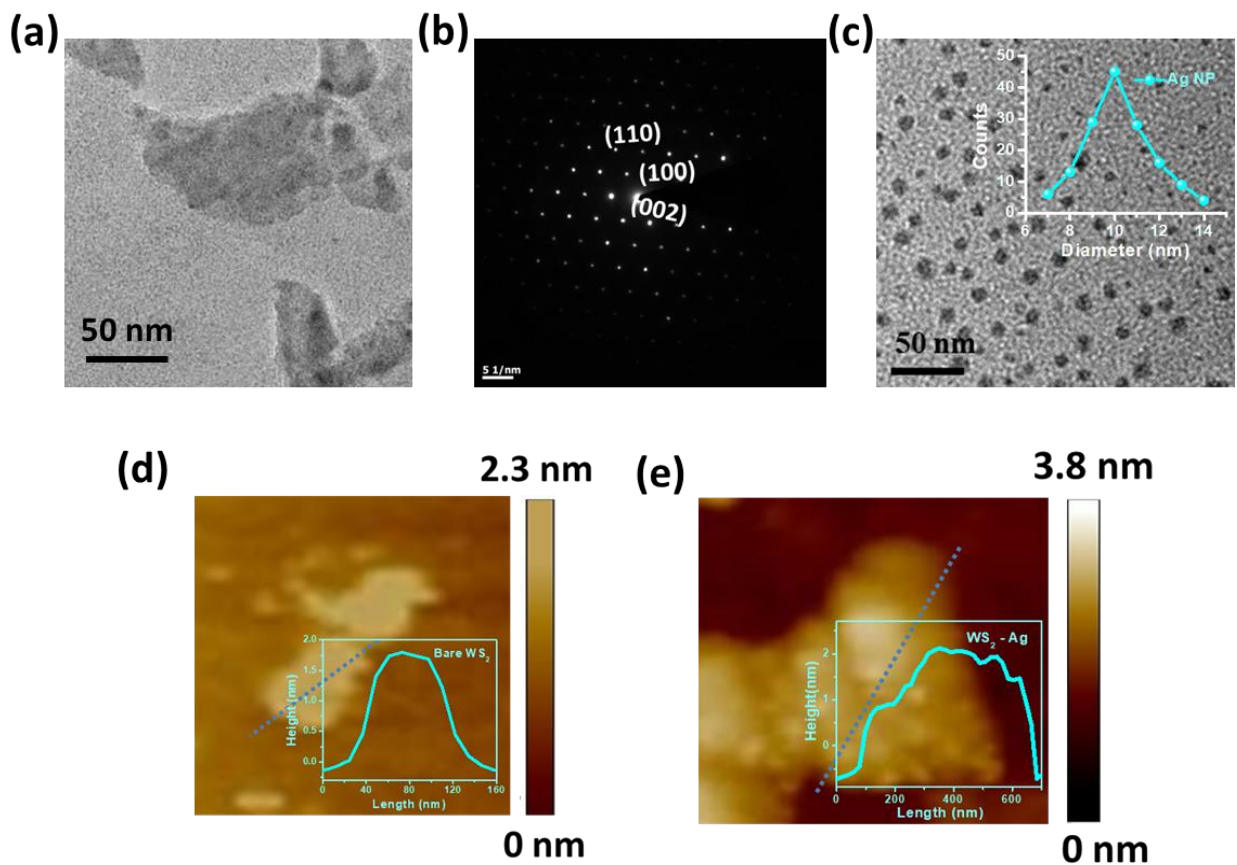

**Figure S1:** Structural characterisation of WS$_2$ layers and Ag nanoparticles (NPs) (a) TEM image and (b) SAED patterns of WS$_2$ nanosheets. (c) TEM image of Ag NPs. Size distribution of Ag NPs (inset) shows the average size of the Ag NPs are 10 nm on diameter. AFM images and the corresponding height profiles (inset)of (d) bare WS$_2$ and (e) WS$_2$-Ag hybrid system respectively.

TEM image of chemically exfoliated WS$_2$ indicates morphology of two dimensional nanosheets (**Figure S1a**). The Selected Area Eletron Diffraction (SAED) indicates (**Figure S1b**) the heagonal lattice structure with good crystallinity of the chemically exfoliated WS$_2$ nanosheets. **Figure S1c** indicates the synthesized silver nanopartiles are almost spherical in nature with the average size distribution of ~10 nm in diameter. The AFM images and the corresponding height profiles (inset)(**Figure S1d** and **S1e**) suggest the thickness of WS$_2$ ~ 2/3 nm corresponds to 3-4 layers in both cases.

## 2. Spectroscopic characterisations of $WS_2$ and the plasmonic $WS_2$-Ag hybrid system:

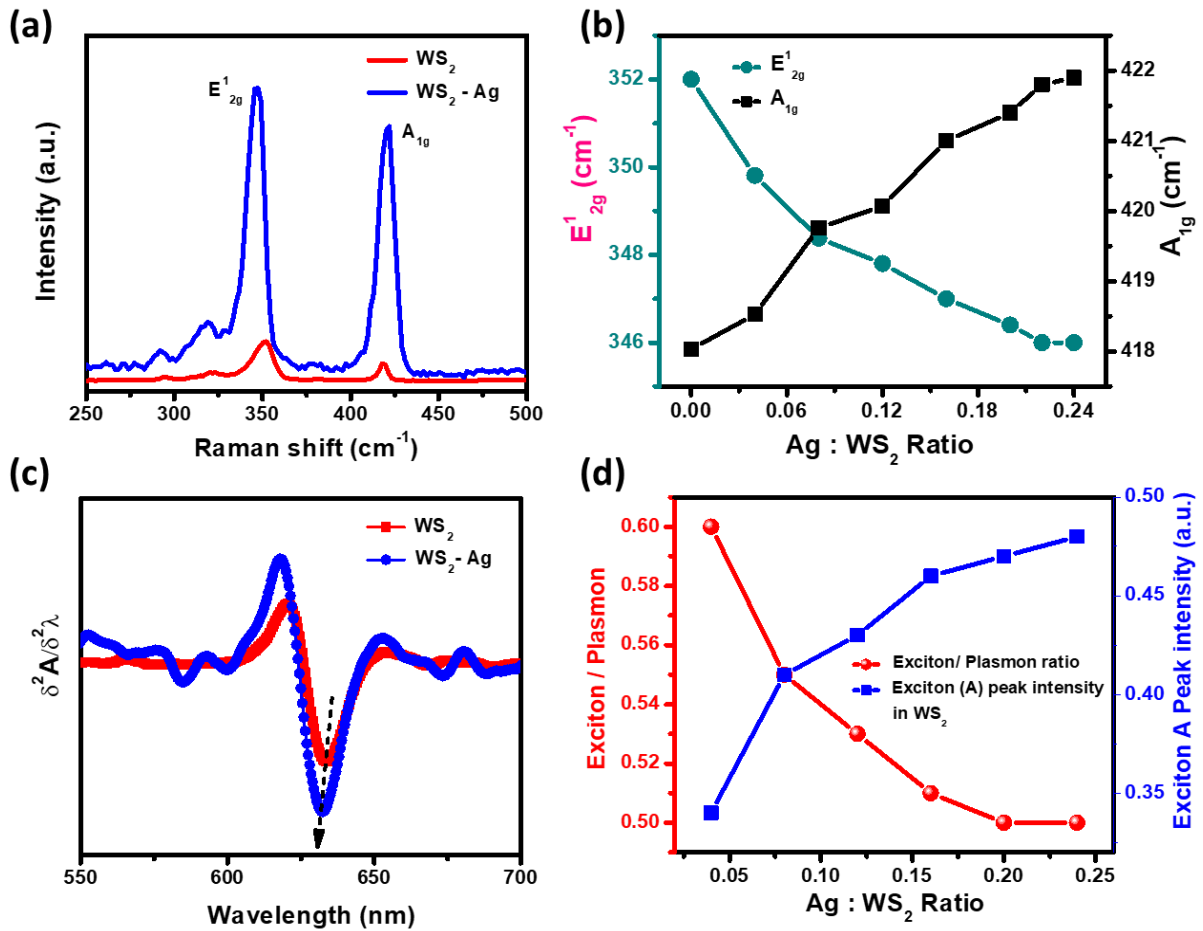

**Figure S2:** Spectroscopic characterisations of plasmonic $WS_2$-Ag hybrid (a) Raman spectra of $WS_2$ and $WS_2$-Ag hybrid system (b) Dependence of in-plane $E^1_{2g}$ and out-of-plane $A_{1g}$ Raman mode with Ag: $WS_2$ ratio. (c) Second derivative on the A exciton region in the absorbance spectra of $WS_2$ and $WS_2$-Ag. (d) Dependence of exciton / plasmon ration and the exciton A peak intensity on $WS_2$: Ag ratio.

## 3. Photoluminescence analysis of WS$_2$ and the plasmonic WS$_2$-Ag hybrid system:

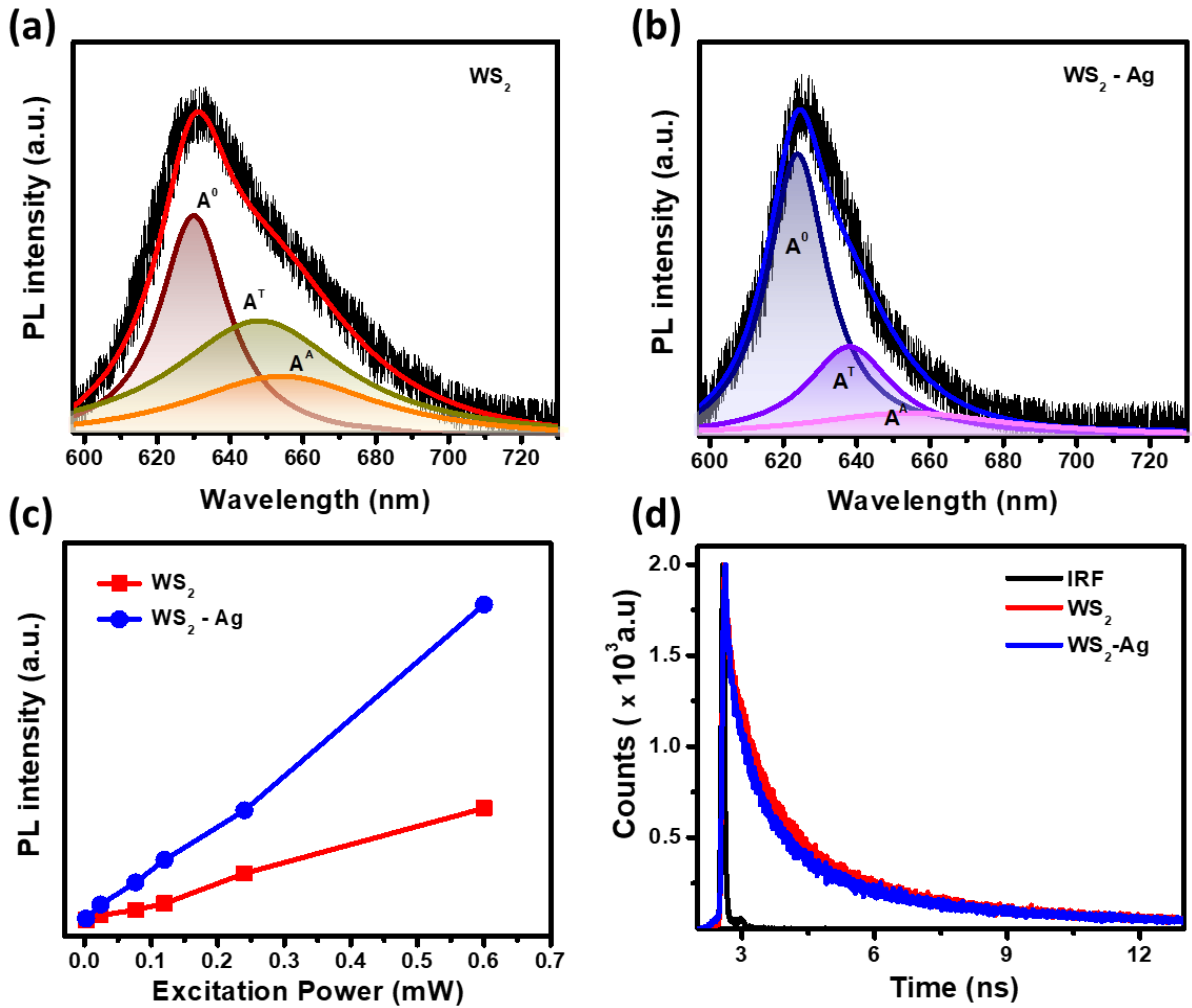

**Figure S3:** Evaluation of photoluminescence (PL) spectra of WS$_2$ and WS$_2$-Ag hybrid (a) and (b) Fitted PL spectra of WS$_2$ and the plasmonic hybrid respectively. The PL spectra for both the systems can be deconvoluted by three distinct species, called neutral exciton (A$^0$), trion (A$^T$) and biexciton (A$^A$). (c) Excitation power dependent PL intensity for bare WS$_2$ and the hybrid system. (d) The time resolved PL (TRPL) decay of the pristine WS$_2$ and the hybrid structure.

# 4. Calculations of the spectral density of noise of the hybrid phototransistors:

The noise equivalent power (NEP) and the specific detectivity ($D^*$) of the phototransistor can be found by considering the total spectral density of noise ($S_I$) of the device in dark current waveform. The spectral density of noise ($S_I$) for a phototransistor is defined as

$$S_I = S_I (Shot) + S_I (Thermal) + S_I (1/f) \quad \ldots\ldots\ldots\ldots (1)$$

The spectral density of shot noise ($S_I$ (Shot)) can be calculated by the equation…

$$S_I (Shot) = 2qI_{dark} \quad \ldots\ldots\ldots\ldots\ldots\ldots\ldots\ldots\ldots (2)$$

Where, q is the electronic charge and $I_{dark}$ is the dark current of the hybrid devices at the same experimental conditions of $V_{ds} = 1$ V and $V_{bg} = 0$ V.

By using the above equation (equation 2), the calculated spectral density of shot noise ($S_I$ (Shot)) of the hybrid graphene/WS$_2$-Ag and graphene/WS$_2$ phototransistors are $2.92 \times 10^{-22}$ A$^2$/Hz and $1.68 \times 10^{-22}$ A$^2$/Hz respectively.

Also, the spectral density of thermal noise ($S_I$(Thermal)) can be calculated by the equation…

$$S_I (Thermal) = \frac{4K_B T}{R} \quad \ldots\ldots\ldots\ldots\ldots\ldots\ldots (6)$$

Where, $K_B$ is the Boltzmann constant, T is the room temperature and R is the device resistance at dark. In room temperature (T = 300K), the $S_I$ (thermal) are calculated to be $1.50 \times 10^{-23}$ A$^2$/Hz and $8.67 \times 10^{-23}$ A$^2$/Hz for the hybrid graphene/WS$_2$-Ag and graphene/WS$_2$ devices respectively.

The 1/f noise spectral density ($S_I$ (1/f)) of the hybrid device is measure directly in dark with $V_{ds} = 1$ V, $V_{bg} = 0$ V (**Figure 4a** and **4b** inset). At modulation frequency 1 Hz, the measured $S_I$ (1/f) is $3.01 \times 10^{-17}$ A$^2$ / Hz for graphene/WS$_2$-Ag and $1.11 \times 10^{-17}$ A$^2$ / Hz for bare for graphene/WS$_2$ hybrid phototransistors.

From these above results it is clear that the 1/f noise dominates over the total current noise spectral density of the hybrid phototransistor device.

# 5. Comparison of temporal photoresponse characteristics of the hybrid devices:

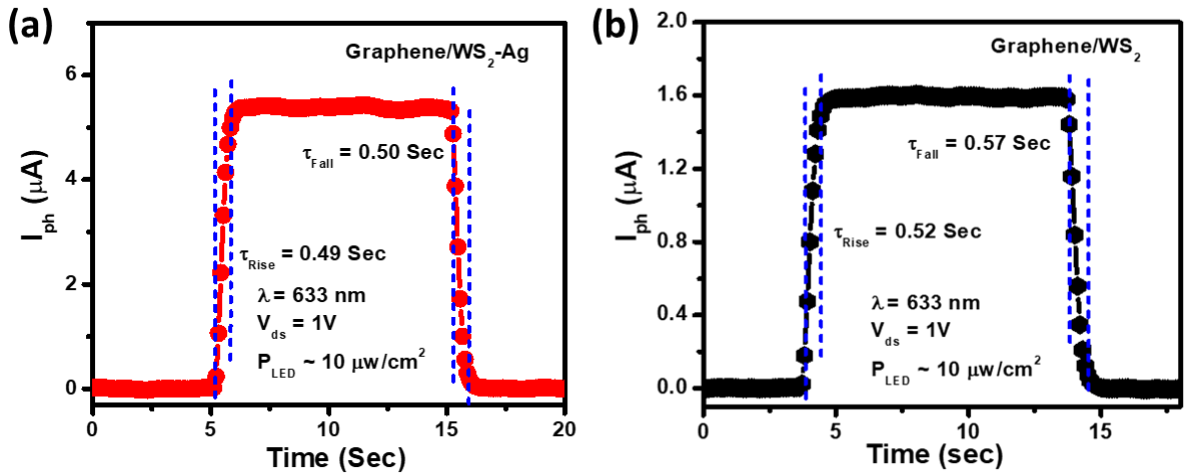

**Figure S4:** The enlarged one cycle view of temporal photoresponse characteristics of hybrid (a) graphene/WS$_2$-Ag and (b) graphene/WS$_2$ phototransistors with λ = 633 nm, $V_{ds}$ = 1V, $V_{bg}$ = 0V, $P_{LED}$ ~ 10μW/ cm$^2$.

# 6. Transfer characteristics and the suggested energy band diagrams:

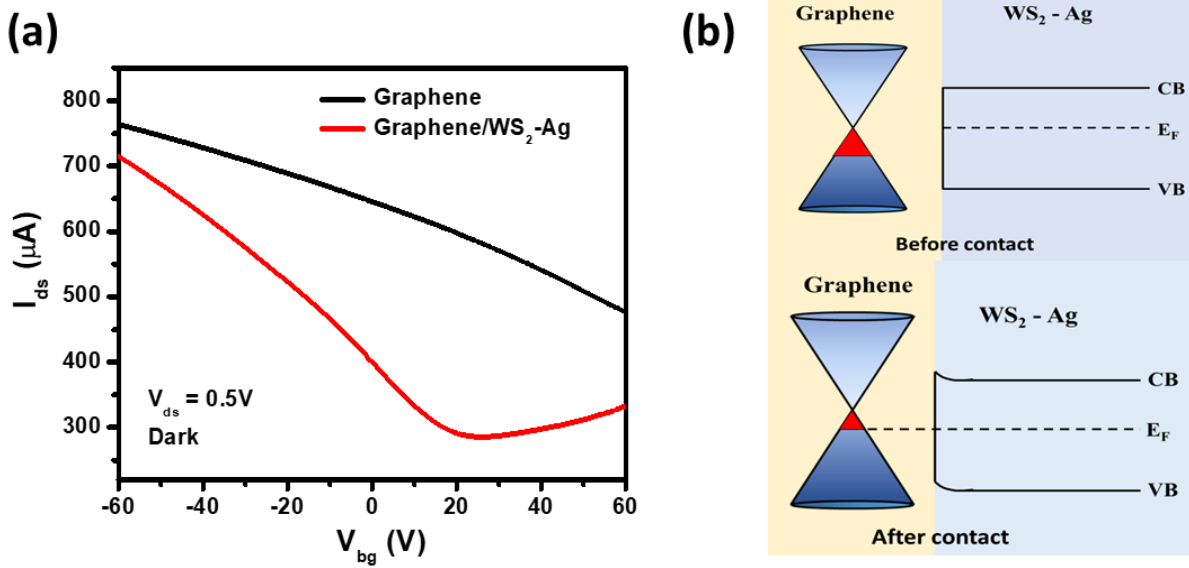

**Figure S5:** (a) Transfer characteristics of the graphene transistor before and after making the heterostructure in dark, $V_{ds}$ = 0.5V. (b) Schematic energy band diagram before (top) and after (bottom) the graphene is contacted with WS$_2$-Ag.

## 7. Stability of the hybrid graphene/WS$_2$-Ag phototransistor:

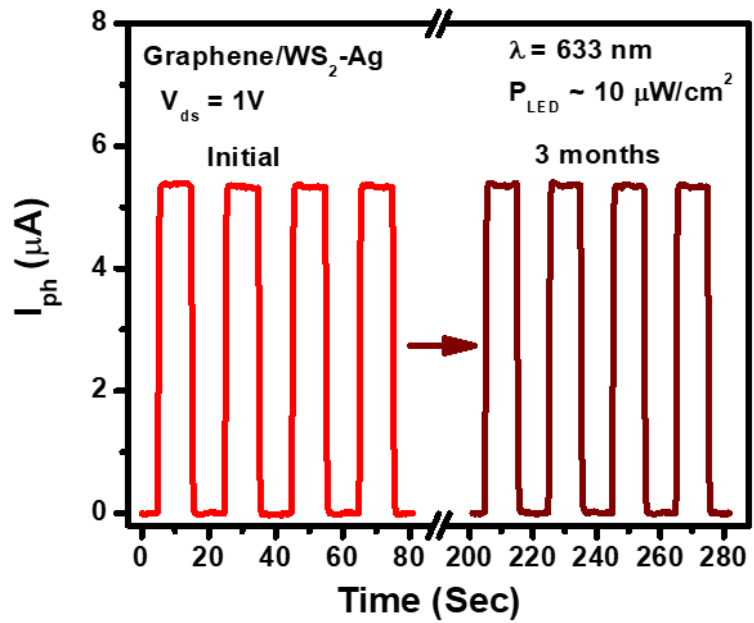

**Figure S6:** The temporal photocurrent (I$_{ph}$) of the graphene/WS$_2$-Ag hybrid phototransistor after 3 months of its fabrication which suggests the excellent stability of the device. The experiments are performed with λ = 633 nm, V$_{ds}$ = 1 V, V$_{bg}$ = 0 V, P$_{LED}$ ~ 10 µW/cm$^2$.